# Density Functional Theory is Not Straying from the Path toward the Exact Functional

Kasper P. Kepp*

*Technical University of Denmark, DTU Chemistry, Building 206, 2800 Kgs. Lyngby, DK – Denmark. Phone: +045 45 25 24 09. * Corresponding e-mail: kpj@kemi.dtu.dk. Submitted Feb 2, 2017*

Abstract

Recently (*Science*, 355, 6320, 2017, 49−52) it was argued that density functionals stray from the path towards exactness due to errors in densities ($\rho$) of 14 atoms and ions computed with several recent functionals. However, this conclusion rests on very compact $\rho$ of highly charged $1s^2$ and $1s^2 2s^2$ systems, the divergence is due to one particular group's recently developed functionals, whereas other recent functionals perform well, and errors in $\rho$ were not compared to actual energies $E[\rho]$ of the same distinct, compact systems, but to general errors for diverse systems. As argued here, a true path can only be defined for $E[\rho]$ and $\rho$ for the *same* systems: By computing errors in $E[\rho]$, it is shown that different functionals show remarkably linear error relationships between $\rho$ and $E[\rho]$ on well-defined but different paths towards exactness, and the ranking in *Science*, 355, 6320, 2017, 49−52 breaks down. For example, M06-2X, said to perform poorly, performs very well on the E,$\rho$ paths defined here, and local (non-GGA) functionals rapidly increase errors in $E[\rho]$ due to the failure to describe dynamic correlation of compact systems without the gradient. Finally, a measure of "exactness" is given by the product of errors in $E[\rho]$ and $\rho$; these relationships may be more relevant focus points than a time line if one wants to estimate exactness and develop new exact functionals.

In their recent paper(*1*), Medvedev et al. point out that electron densities $\rho$ and energies $E[\rho]$ computed with density functional theory (DFT) not always increase in accuracy together. Burke et al.(*2*) stated the problem in 1998 as "functionals which yield highly accurate energies often produce potentials which differ markedly from the exact ones." Medvedev et al. put errors in $\rho$ on a time scale and show a trend of improvement impaired by nine specific recent functionals mostly using a high-parameterization philosophy with reported high accuracy of $E[\rho]$ for diverse systems. The inverse relationship in Medvedev et al. may suggest an overfitting problem on the path towards universality, where both $\rho$ and $E[\rho]$ should become increasingly accurate; off this track, accurate energies with inaccurate densities would seem successful only until applied outside the parameterization range. Some comments seem warranted:

1) Of the 9 specific functionals that deviate from the "path", almost all are from 2011-2012 and all from one specific group; other recent functionals perform well in the trend, and the only two functionals from 2014/2015 are on-path. With two functionals from 2015, none from 2014, and three from 2013, the recent history seems under-sampled; various post-2011 functionals by other groups that would define the trend of have been not included(*3*)(*4*)(*5*)(*6*)(*7*). Thus, whereas *some* recent functionals from one research group have sacrificed accuracy in $1s^2$ and $1s^2 2s^2$ atomic ion densities for accuracy in diverse molecular energies, arguing that DFT recently deviates from the path seems too generalizing.

2) A concern is whether the functionals are actually on a "path" as no direct comparison of E[$\rho$] and $\rho$ was done; the errors in E were from general benchmarks of diverse molecules(*8*). There is only a path if the errors of both E[$\rho$] and $\rho$ decrease together for the *same* systems; and the $\rho$ of the studied systems are distinctly different from those of typical systems (*vide infra*).

3) The errors in $\rho$ grow roughly with $\sqrt{n}$ (n = the number of electrons)(see e.g. File S3 of Medvedev et al.). If one divides each error with $\sqrt{n}$, the standard deviation in error for different *n*-electron systems falls from 0.68 to 0.14, and the scaled errors are not significantly different between systems, as expected for *n* random variables with an independent, constant error that relates to the failure of producing the electron pair correlation.

4) HF recovers most of the correlation energy of the $1s^2$ systems ($B^{3+}$, $C^{4+}$, $N^{5+}$, $O^{6+}$, $F^{7+}$, and $Ne^{8+}$) ; its RMSD for $\rho$ is only 0.049 for these systems (File S3 of Medvedev et al.). Accordingly, functionals with HF exchange perform more accurately for the $1s^2$ systems. 6 of the 14 systems studied (43%) are of this type. The top-performers are therefore hybrid functionals that fit the benchmark systems. However, the high accuracy of the HF picture is unique for systems with $2N^2$ valence electrons (the octet rule), where N is the period number (e.g. Ne requires much more HF exchange, as do the $1s^2$ systems). If one leaves out the six 2-electron systems, HF performs poorly (average RMSD of $\rho$ = 1.81 without 2-electron systems, 0.92 with).

5) Similarly, Medvedev et al. report in a figure the maximum combined error of $\rho$, its gradient, and Laplacian (one can discuss the relevance of the latter); including the six $1s^2$ systems would reveal the high HF demands of the $1s^2$ configurations, and for the major part of periodic table, smaller HF percentages are required(*9*)(*10*), so the appraisal of 25% HF exchange is specific to $1s^22s^2$ systems where the gap between virtual and occupied orbitals justifies 25%. Thus, a figure with all systems included would have shown that the HF percentage required is system-dependent and there is no magic 25%.

6) 13 of the 14 systems are $1s^2$ or $1s^22s^2$ systems and 10 of the 14 studied ions have a charge between +3 and +8, representing extremely compact $\rho$ with large dynamic correlation, viz. the large improvement by MP4 over MP2 (File S1 of Medvedev et al.). For real molecules, localized charges of +3 are not seen because charge delocalizes onto neighbor atoms. Thus, while the errors in $\rho$ are notable, it is unclear if the deviation from the exact $\rho$ near the nucleus of a very compact density is chemically relevant.

To address 1–6 in a combined way, because energy is a state function, the quality of E[$\rho$] for a given functional can be probed by comparing to ionization potentials (IP) from the NIST data base, e.g. E[$\rho$] of $B^{3+}$ and $B^+$ can be probed by the 2nd and 3rd experimental IP of boron (di-cation energies cancel out);

$$E(B^{3+}) - E(B^+) = IP3(B) + IP2(B) = 37.931 \text{ eV} + 25.155 \text{ eV} = 63.085 \text{ eV} \qquad (1)$$

These energies correspond to removing both 2s electrons from the $1s^22s^2$ configurations, with a trend of increasing charge and more compact $\rho$. Comparing to E[$\rho$] directly reveals 1) whether the reported errors in $\rho$ have chemical relevance on the energy scale, 2) whether there is a relationship between errors E[$\rho$] and $\rho$ implying a "path" towards universality, and accordingly, a deviation from such path, as claimed.

Computations where carried out using the software Turbomole 7.0(*11*) for $E(B^{3+}) - E(B^+) = IP3(B) + IP2(B) = 63.085$ eV, $E(C^{4+}) - E(C^{2+}) = IP4(C) + IP3(C) = 112.381$ eV, $E(N^{5+}) - E(N^{3+}) = IP5(N) + IP4(N) = 175.364$ eV, $E(O^{6+}) - E(O^{4+}) = IP6(O) + IP5(O) = 252.018$ eV, $E(F^{7+}) - E(F^{5+}) = IP7(F) + IP6(F) = 342.350$ eV, and $E(Ne^{8+}) - E(Ne^{6+}) = IP8(Ne) + IP7(Ne) = 446.368$ eV. This benchmark of covers 12 of the 14 systems studied by Medvedev et al.

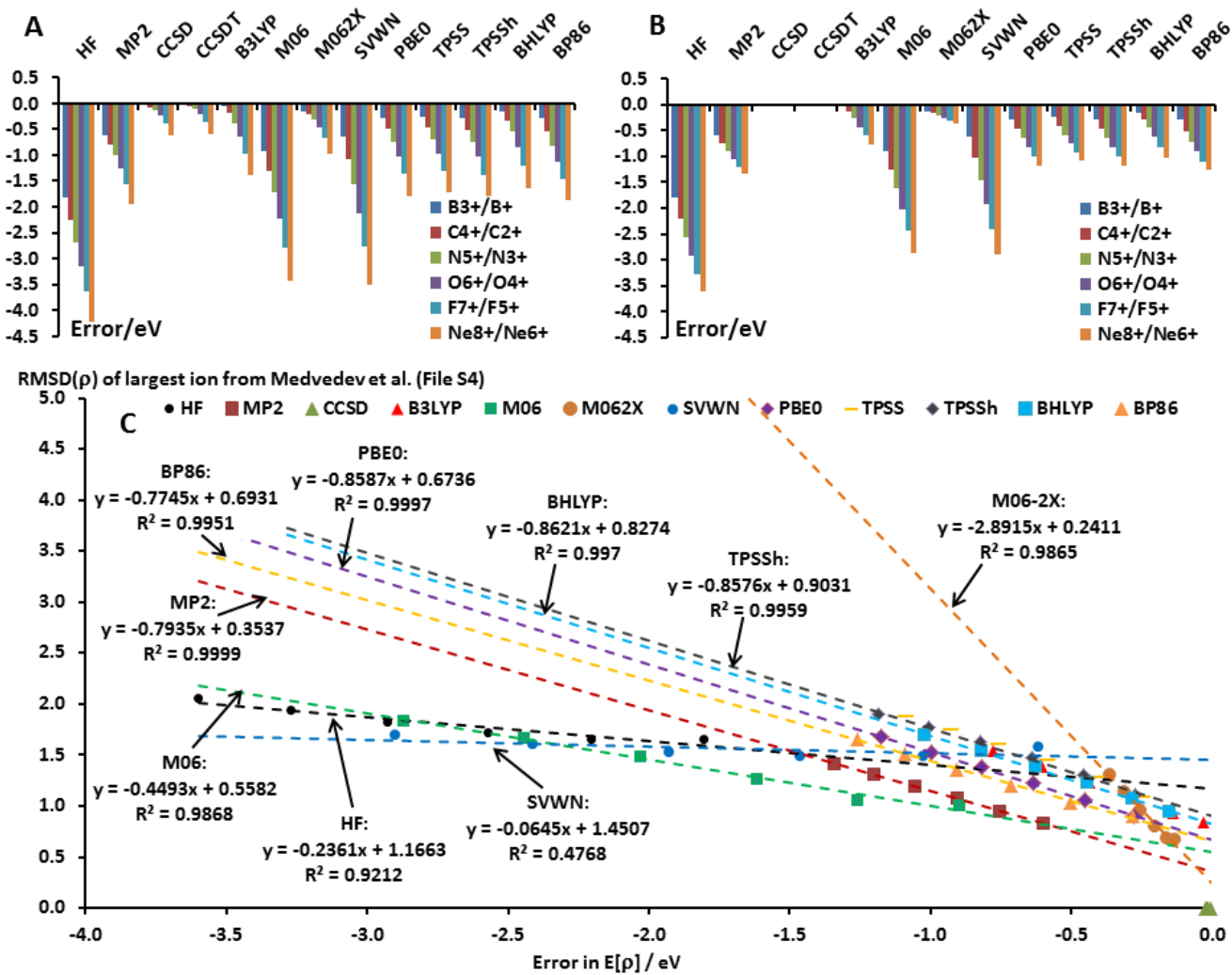

**Figure 1. Paths of Accuracy: A) Non-relativistic and B) relativistic errors in computed ionic energy differences vs. experimental, in eV (equation 1). C) Errors in densities of larger ion vs. errors in computed energies.**

To ensure stringent comparison, the same aug-cc-pwCV5Z basis set was used, and densities and energies were converged to $10^{-7}$ and $10^{-8}$ a.u. using ultra-fine grids (m5). To illustrate the general features of such paths, PBE0, TPSSh, and TPSS were chosen as non-empirical functionals, B3LYP as a commonly used functional and BHLYP as its well-performing (in Medvedev et al.) half-and half HF version, BP86 as a classical GGA, M06 and SVWN as local functionals, M06-2X as a low-ranked (in Medvedev et al.) empirical functional of the Minnesota type(*12*), in addition to HF, MP2, and CCSD. CCSD(T) was also calculated, as CCSD is full-CI and thus exact non-relativistic for $1s^2$ systems, but may miss some core-valence correlation of the 4- and 10-electron systems.

The results in **Figure 1A** (non-relativistic) and **Figure 1B** (relativistic corrected) show that relativistic effects grow with the charge, as 1s-electrons are accelerated (numerical values are shown in Appendix **Table 1** and **Table 2**). Relativistic stabilization and contraction of the s-shells favor the $1s^2 2s^2$ systems over $1s^2$ systems. Due to zero spin and angular momentum, scalar relativistic corrections recover this effect effectively (**Figure 1B**) and are quite large for the highly charged ions, >0.6 eV for the neon systems (the neon-systems have the largest errors in Medvedev et al., as probably even ρ is affected by relativistic s-shell contraction).

Relativistic corrected CCSD(T) and CCSD energies are within 0.03 eV (~3 kJ/mol) of experiment, because the strong dynamic correlation is well described. Accordingly, the exact density functional methodology would provide exact energies to within 3 kJ/mol if applied with this basis set and relativistic correction. Thus, we can compare the density functionals now also in the energy regime, $E[\rho]$.

HF energies show errors exceeding 3 eV for neon systems (**Figure 1B**). Local functionals M06 and SVWN that performed poorly for $\rho$ produce energy errors almost as large as HF. Most other functionals perform similarly although B3LYP and M06-2X perform distinctly better. Also, BP86 performs similar to functionals such as TPSSh and TPSS that scored highly in Medvedev et al. due to exact constraints that improve their core density(*13*).

In **Figure 1B**, only the first bar represents a realistic $\rho$. Net atomic charges rarely exceed 1 even for highly charged groups such as phosphates and high-valent metal sites. For the chemically relevant boron densities all DFT methods perform better than MP2, which only becomes more accurate as the dynamical correlation increases in the extremely compact density limit of highly charged ions. The error of B3LYP is 0.03 eV, and the worst performing (PBE0, BP86) is 0.27–0.28 eV. These errors are typical of chemically relevant energies; thus, the extremely compact regime mostly studied by Medvedev et al. is probably not chemically relevant.

Instead, in order to compare same-system energies and densities as required by a well-defined path towards exactness, RMSD values of $\rho$ for the largest $1s^2 2s^2$ ions from Medvedev et al. (File S4) are compared to error in energy of removing the two $2s^2$ electrons. **Figure 1C** reveals strong linear relationships: Since all the energies are for the same iso-electronic conversions ($1s^2 2s^2$ systems where the 2s electrons are removed) they reflect monotonous but distinct trends in sensitivity to increased charge, which increases kinetic energy and dynamic correlation as $\rho$ becomes more compact. "Exactness" is represented by CCSD(T) in the right lower corner.

Most DFT methods and MP2 follow the same "path" of accuracy with errors in energy growing with errors in $\rho$ (coefficients of −0.77 to −0.86). The local functional SVWN and HF show less linear behavior. M06-2X errors in $E[\rho]$ increase slowly with $\rho$, whereas for local functionals and HF they energies deteriorate much more rapidly as $\rho$ becoems denser because they do not handle dynamic correlation well in this limit due to not having the gradient included. From this comparison of the density and energy regime, M06-2X is the most "exact" functional for these systems, and much more exact than MP2, PBE0, or TPSSh. In Medvedev et al. M06-2X is ranked low mainly because of gradients and Laplacian of $\rho$ and thus, for this reason claimed to be off path, despite $E[\rho]$ and $\rho$ being excellently on path (**Figure 1C**).

It is necessary to quantify exactness on a path of *both* $E[\rho]$ and $\rho$, since wrong densities can give right energies, and right densities can give wrong energies. To define a measure of "exactness" one can therefore use the area of the rectangle defined by a given point of **Figure 1C**. For the most challenging dense cases, these measures of exactness are: 0 (CCSD/CCSD(T)), 0.5 (M06-2X), 1.2 (B3LYP), 1.7 (BHLYP), 1.9 (MP2), 2.0 (PBE0, TPSS), 2.1 (BP86), 2.3 (TPSSh), 4.9 (SVWN), 5.3 (M06), and 7.4 (HF), a ranking very different from that of Medvedev et al.

In conclusion, the stated poor performance of some recent functionals for very compact $\rho$ does not imply that they are less exact, partly because E and $\rho$ were not compared for same systems, and partly because hybrid functionals are favored by the choice of benchmark systems. Instead, paths are defined here of both E and $\rho$. All functionals show linear E,$\rho$ paths, and different functional types show distinct error relationships between $\rho$ and $E[\rho]$. These relationships are on actual, but different paths likely to be of interest if one wants to produce exact functionals; a measure of exactness is suggested for this purpose.

**Appendix Table 1. Errors in computed energies according to Equation (1), in eV.**

NON-RELATIVISTIC

|  | HF | MP2 | CCSD | CCSDT | B3LYP | M06 | M062X |
|---|---|---|---|---|---|---|---|
| $B^{3+}-B^{+}$ | -1.8239 | -0.6133 | -0.0394 | -0.0188 | -0.0468 | -0.9132 | -0.1485 |
| $C^{4+}-C^{2+}$ | -2.2511 | -0.7987 | -0.0693 | -0.0467 | -0.1845 | -1.2994 | -0.2061 |
| $N^{5+}-N^{3+}$ | -2.6790 | -1.0079 | -0.1238 | -0.1001 | -0.3804 | -1.7299 | -0.3070 |
| $O^{6+}-O^{4+}$ | -3.1389 | -1.2635 | -0.2248 | -0.2001 | -0.6430 | -2.2285 | -0.4588 |
| $F^{7+}-F^{5+}$ | -3.6443 | -1.5735 | -0.3805 | -0.3551 | -0.9772 | -2.7915 | -0.6818 |
| $Ne^{8+}-Ne^{6+}$ | -4.2147 | -1.9548 | -0.6074 | -0.5816 | -1.3943 | -3.4430 | -0.9842 |

|  | SVWN | PBE0 | TPSS | TPSSh | BHLYP | BP86 |
|---|---|---|---|---|---|---|
| $B^{3+}-B^{+}$ | -0.6332 | -0.2928 | -0.2603 | -0.2920 | -0.1675 | -0.2981 |
| $C^{4+}-C^{2+}$ | -1.0722 | -0.5017 | -0.4598 | -0.5042 | -0.3293 | -0.5527 |
| $N^{5+}-N^{3+}$ | -1.5678 | -0.7407 | -0.6889 | -0.7461 | -0.5499 | -0.8208 |
| $O^{6+}-O^{4+}$ | -2.1367 | -1.0275 | -0.9666 | -1.0367 | -0.8374 | -1.1210 |
| $F^{7+}-F^{5+}$ | -2.7781 | -1.3722 | -1.3008 | -1.3839 | -1.1957 | -1.4674 |
| $Ne^{8+}-Ne^{6+}$ | -3.5055 | -1.7898 | -1.7080 | -1.8041 | -1.6363 | -1.8766 |

RELATIVISTIC

|  | HF | MP2 | CCSD | CCSDT | B3LYP | M06 | M062X |
|---|---|---|---|---|---|---|---|
| $B^{3+}-B^{+}$ | -1.8069 | -0.5963 | **-0.0224** | **-0.0018** | -0.0294 | -0.8984 | -0.1327 |
| $C^{4+}-C^{2+}$ | -2.2035 | -0.7511 | **-0.0217** | **0.0008** | -0.1364 | -1.2580 | -0.1597 |
| $N^{5+}-N^{3+}$ | -2.5723 | -0.9011 | **-0.0171** | **0.0067** | -0.2726 | -1.6175 | -0.2027 |
| $O^{6+}-O^{4+}$ | -2.9283 | -1.0529 | **-0.0142** | **0.0104** | -0.4311 | -2.0317 | -0.2527 |
| $F^{7+}-F^{5+}$ | -3.2704 | -1.1996 | **-0.0066** | **0.0188** | -0.6014 | -2.4433 | -0.3097 |
| $Ne^{8+}-Ne^{6+}$ | -3.5996 | -1.3396 | **0.0077** | **0.0335** | -0.7770 | -2.8699 | -0.3618 |

|  | SVWN | PBE0 | TPSS | TPSSh | BHLYP | BP86 |
|---|---|---|---|---|---|---|
| $B^{3+}-B^{+}$ | -0.6164 | -0.2755 | -0.2431 | -0.2748 | -0.1502 | -0.2808 |
| $C^{4+}-C^{2+}$ | -1.0253 | -0.4536 | -0.4117 | -0.4562 | -0.2813 | -0.5047 |
| $N^{5+}-N^{3+}$ | -1.4625 | -0.6332 | -0.5813 | -0.6385 | -0.4423 | -0.7132 |
| $O^{6+}-O^{4+}$ | -1.9290 | -0.8160 | -0.7548 | -0.8250 | -0.6258 | -0.9092 |
| $F^{7+}-F^{5+}$ | -2.4089 | -0.9968 | -0.9243 | -1.0077 | -0.8204 | -1.0915 |
| $Ne^{8+}-Ne^{6+}$ | -2.8976 | -1.1730 | -1.0882 | -1.1850 | -1.0196 | -1.2587 |

**Appendix Table 2. Electronic energies of computed systems (in a.u.) and errors vs. experiment (in eV).**

| EXP/eV | HF | scalar rel | TOTAL |  | MP2 | scalar rel | TOTAL |
|---|---|---|---|---|---|---|---|
| $B^{3+}$ | -21.9862297 | -0.0062292 | -21.9924589 |  | -22.0276130 | -0.0062292 | -22.0338422 |
| $B^{+}$ | -24.2375477 | -0.0068538 | -24.2444015 |  | -24.3234197 | -0.0068538 | -24.3302735 |
| $B^{3+}-B^{+}$ | 61.2614823 |  | 61.2784789 |  | 62.4720823 |  | 62.4890789 |
| ERROR/eV | -1.8239 |  | -1.8069 |  | -0.6133 |  | -0.5963 |
|  |  |  |  |  |  |  |  |
| $C^{4+}$ | -32.3612111 | -0.0135718 | -32.3747829 |  | -32.4032075 | -0.0135718 | -32.4167793 |
| $C^{2+}$ | -36.4084214 | -0.0153189 | -36.4237403 |  | -36.5037928 | -0.0153189 | -36.5191117 |
| $C^{4+}-C^{2+}$ | 110.1302023 |  | 110.1777427 |  | 111.5826094 |  | 111.6301498 |
| ERROR/eV | -2.2511 |  | -2.2035 |  | -0.7987 |  | -0.7511 |
|  |  |  |  |  |  |  |  |
| $N^{5+}$ | -44.7361414 | -0.0259800 | -44.7621214 |  | -44.7785775 | -0.0259800 | -44.8045575 |
| $N^{3+}$ | -51.0821835 | -0.0299038 | -51.1120873 |  | -51.1860333 | -0.0299038 | -51.2159371 |
| $N^{5+}-N^{3+}$ | 172.6846000 |  | 172.7913737 |  | 174.3557510 |  | 174.4625246 |
| ERROR/eV | -2.6790 |  | -2.5723 |  | -1.0079 |  | -0.9011 |
|  |  |  |  |  |  |  |  |
| $O^{6+}$ | -59.1114319 | -0.04542472 | -59.1568566 |  | -59.1541848 | -0.04542472 | -59.1996095 |
| $O^{4+}$ | -68.2575633 | -0.0531622 | -68.3107255 |  | -68.3692364 | -0.0531622 | -68.4223986 |
| $O^{6+}-O^{4+}$ | 248.8789147 |  | 249.0894622 |  | 250.7543244 |  | 250.9648720 |
| ERROR/eV | -3.1389 |  | -2.9283 |  | -1.2635 |  | -1.0529 |
|  |  |  |  |  |  |  |  |
| $F^{7+}$ | -75.4866735 | -0.0743298 | -75.5610033 |  | -75.5296969 | -0.0743298 | -75.6040266 |
| $F^{5+}$ | -87.9338732 | -0.0880702 | -88.0219434 |  | -88.0529999 | -0.0880702 | -88.1410701 |
| $F^{7+}-F^{5+}$ | 338.7055536 |  | 339.0794500 |  | 340.7764329 |  | 341.1503293 |
| ERROR/eV | -3.6443 |  | -3.2704 |  | -1.5735 |  | -1.1996 |
|  |  |  |  |  |  |  |  |
| $Ne^{8+}$ | -93.8619521 | -0.1152036 | -93.9771557 |  | -93.9051977 | -0.1152036 | -94.0204013 |
| $Ne^{6+}$ | -110.1107825 | -0.1378085 | -110.2485910 |  | -110.2370799 | -0.1378085 | -110.3748884 |
| $Ne^{8+}-Ne^{6+}$ | 442.1531957 |  | 442.7683087 |  | 444.4131489 |  | 445.0282618 |
| ERROR/eV | -4.2147 |  | -3.5996 |  | -1.9548 |  | -1.3396 |

| EXP/eV | CCSD | scalar rel | TOTAL | | CCSD(T) | scalar rel | TOTAL |
|---|---|---|---|---|---|---|---|
| $B^{3+}$ | -22.0299257 | -0.0062292 | -22.0361549 | | -22.0299257 | -0.0062292 | -22.0361549 |
| $B^{+}$ | -24.3468250 | -0.0068538 | -24.3536788 | | -24.3475804 | -0.0068538 | -24.3544343 |
| $B^{3+}-B^{+}$ | 63.0460402 | | 63.0630368 | | 63.0665977 | | 63.0835943 |
| ERROR/eV | -0.0394 | | -0.0224 | | -0.0188 | | -0.0018 |
| | | | | | | | |
| $C^{4+}$ | -32.4051589 | -0.0135718 | -32.4187307 | | -32.4051589 | -0.0135718 | -32.4187307 |
| $C^{2+}$ | -36.5325497 | -0.0153189 | -36.5478686 | | -36.5333773 | -0.0153189 | -36.5486962 |
| $C^{4+}-C^{2+}$ | 112.3120258 | | 112.3595662 | | 112.3345461 | | 112.3820864 |
| ERROR/eV | -0.0693 | | -0.0217 | | -0.0467 | | 0.0008 |
| | | | | | | | |
| $N^{5+}$ | -44.7802664 | -0.0259800 | -44.8062464 | | -44.7802664 | -0.0259800 | -44.8062464 |
| $N^{3+}$ | -51.2202101 | -0.0299038 | -51.2501139 | | -51.2210841 | -0.0299038 | -51.2509879 |
| $N^{5+}-N^{3+}$ | 175.2397927 | | 175.3465663 | | 175.2635761 | | 175.3703498 |
| ERROR/eV | -0.1238 | | -0.0171 | | -0.1001 | | 0.0067 |
| | | | | | | | |
| $O^{6+}$ | -59.1556758 | -0.04542472 | -59.2011006 | | -59.1556758 | -0.04542472 | -59.2011006 |
| $O^{4+}$ | -68.4088985 | -0.0531622 | -68.4620607 | | -68.4098051 | -0.0531622 | -68.4629673 |
| $O^{6+}-O^{4+}$ | 251.7930144 | | 252.0035619 | | 251.8176824 | | 252.0282299 |
| ERROR/eV | -0.2248 | | -0.0142 | | -0.2001 | | 0.0104 |
| | | | | | | | |
| $F^{7+}$ | -75.5310297 | -0.0743298 | -75.6053594 | | -75.5310297 | -0.0743298 | -75.6053594 |
| $F^{5+}$ | -88.0981751 | -0.0880702 | -88.1862453 | | -88.0991059 | -0.0880702 | -88.1871761 |
| $F^{7+}-F^{5+}$ | 341.9694443 | | 342.3433407 | | 341.9947745 | | 342.3686709 |
| ERROR/eV | -0.3805 | | -0.0066 | | -0.3551 | | 0.0188 |
| | | | | | | | |
| $Ne^{8+}$ | -93.9064023 | -0.1152036 | -94.0216059 | | -93.9064023 | -0.1152036 | -94.0216059 |
| $Ne^{6+}$ | -110.2877982 | -0.1378085 | -110.4256067 | | -110.2887480 | -0.1378085 | -110.4265566 |
| $Ne^{8+}-Ne^{6+}$ | 445.7604852 | | 446.3755982 | | 445.7863327 | | 446.4014457 |
| ERROR/eV | -0.6074 | | 0.0077 | | -0.5816 | | 0.0335 |

| EXP/eV | B3LYP | scalar rel | TOTAL | | M06 | scalar rel | TOTAL |
|---|---|---|---|---|---|---|---|
| $B^{3+}$ | -22.0131742 | -0.0064922 | -22.0196664 | | -22.0491882 | -0.0063438 | -22.0555319 |
| $B^{+}$ | -24.3297999 | -0.0071306 | -24.3369305 | | -24.3339733 | -0.0068893 | -24.3408626 |
| $B^{3+}-B^{+}$ | 63.0385944 | | 63.0559666 | | 62.1721694 | | 62.1870140 |
| ERROR/eV | -0.0468 | | -0.0294 | | -0.9132 | | -0.8984 |
| | | | | | | | |
| $C^{4+}$ | -32.3791998 | -0.0140260 | -32.3932258 | | -32.4265191 | -0.0138544 | -32.4403736 |
| $C^{2+}$ | -36.5023552 | -0.0157951 | -36.5181503 | | -36.5087050 | -0.0153753 | -36.5240803 |
| $C^{4+}-C^{2+}$ | 112.1967742 | | 112.2449129 | | 111.0819360 | | 111.1233196 |
| ERROR/eV | -0.1845 | | -0.1364 | | -1.2994 | | -1.2580 |
| | | | | | | | |
| $N^{5+}$ | -44.7447832 | -0.0266901 | -44.7714733 | | -44.8033591 | -0.0259800 | -44.8293390 |
| $N^{3+}$ | -51.1752997 | -0.0306519 | -51.2059516 | | -51.1842793 | -0.0301122 | -51.2143915 |
| $N^{5+}-N^{3+}$ | 174.9832663 | | 175.0910719 | | 173.6336833 | | 173.7461268 |
| ERROR/eV | -0.3804 | | -0.2726 | | -1.7299 | | -1.6175 |
| | | | | | | | |
| $O^{6+}$ | -59.1104082 | -0.04648755 | -59.1568958 | | -59.1802071 | -0.04632277 | -59.2265299 |
| $O^{4+}$ | -68.3482627 | -0.0542752 | -68.4025378 | | -68.3597939 | -0.05355563 | -68.4133496 |
| $O^{6+}-O^{4+}$ | 251.3748227 | | 251.5867352 | | 249.7892799 | | 249.9860958 |
| ERROR/eV | -0.6430 | | -0.4311 | | -2.2285 | | -2.0317 |
| | | | | | | | |
| $F^{7+}$ | -75.4758177 | -0.0758403 | -75.5516580 | | -75.5567322 | -0.0757272 | -75.6324595 |
| $F^{5+}$ | -88.0210349 | -0.0896493 | -88.1106842 | | -88.0352743 | -0.0885220 | -88.1237963 |
| $F^{7+}-F^{5+}$ | 341.3727475 | | 341.7485077 | | 339.5584253 | | 339.9065895 |
| ERROR/eV | -0.9772 | | -0.6014 | | -2.7915 | | -2.4433 |
| | | | | | | | |
| $Ne^{8+}$ | -93.8411225 | -0.1172747 | -93.9583972 | | -93.9330601 | -0.1171940 | -94.0502541 |
| $Ne^{6+}$ | -110.1936005 | -0.1399609 | -110.3335614 | | -110.2102509 | -0.1382525 | -110.3485034 |
| $Ne^{8+}-Ne^{6+}$ | 444.9735916 | | 445.5909156 | | 442.9249226 | | 443.4979524 |
| ERROR/eV | -1.3943 | | -0.7770 | | -3.4430 | | -2.8699 |

| EXP/eV | M06-2X | scalar rel | TOTAL | | SVWN | scalar rel | TOTAL |
|---|---|---|---|---|---|---|---|
| $B^{3+}$ | -22.0342083 | -0.00673916 | -22.0409475 | | -21.7431676 | -0.00614385 | -21.7493114 |
| $B^{+}$ | -24.3470959 | -0.00732138 | -24.3544173 | | -24.038243 | -0.00676041 | -24.0450034 |
| $B^{3+}-B^{+}$ | 62.9368773 | | 62.9527203 | | 62.4521849 | | 62.4689623 |
| ERROR/eV | -0.1485 | | -0.1327 | | -0.6332 | | -0.6164 |
| | | | | | | | |
| $C^{4+}$ | -32.4083416 | -0.01443317 | -32.4227748 | | -32.0394866 | -0.01339684 | -32.0528834 |
| $C^{2+}$ | -36.5307030 | -0.01613771 | -36.5468408 | | -36.1300208 | -0.01511829 | -36.1451391 |
| $C^{4+}-C^{2+}$ | 112.1751681 | | 112.2215507 | | 111.3091072 | | 111.3559504 |
| ERROR/eV | -0.2061 | | -0.1597 | | -1.0722 | | -1.0253 |
| | | | | | | | |
| $N^{5+}$ | -44.7819062 | -0.02733069 | -44.8092369 | | -44.3342002 | -0.02567674 | -44.3598769 |
| $N^{3+}$ | -51.2151196 | -0.0311615 | -51.2462811 | | -50.7210797 | -0.02954603 | -50.7506257 |
| $N^{5+}-N^{3+}$ | 175.0566536 | | 175.1608952 | | 173.7958430 | | 173.9011319 |
| ERROR/eV | -0.3070 | | -0.2027 | | -1.5678 | | -1.4625 |
| | | | | | | | |
| $O^{6+}$ | -59.1553822 | -0.0474746 | -59.2028569 | | -58.6282191 | -0.04492744 | -58.6731465 |
| $O^{4+}$ | -68.4000067 | -0.05504682 | -68.4550536 | | -67.8111813 | -0.05255971 | -67.8637410 |
| $O^{6+}-O^{4+}$ | 251.5590461 | | 251.7650956 | | 249.8811297 | | 250.0888143 |
| ERROR/eV | -0.4588 | | -0.2527 | | -2.1367 | | -1.9290 |
| | | | | | | | |
| $F^{7+}$ | -75.5285563 | -0.07722242 | -75.6057787 | | -74.9212867 | -0.07358192 | -74.9948687 |
| $F^{5+}$ | -88.0846291 | -0.09089496 | -88.1755240 | | -87.4003204 | -0.08715076 | -87.4874711 |
| $F^{7+}-F^{5+}$ | 341.6681443 | | 342.0401930 | | 339.5718016 | | 339.9410284 |
| ERROR/eV | -0.6818 | | -0.3097 | | -2.7781 | | -2.4089 |
| | | | | | | | |
| $Ne^{8+}$ | -93.9015361 | -0.11924706 | -94.0207832 | | -93.2136971 | -0.11412856 | -93.3278256 |
| $Ne^{6+}$ | -110.269085 | -0.14212043 | -110.4112050 | | -109.488588 | -0.13646911 | -109.6250575 |
| $Ne^{8+}-Ne^{6+}$ | 445.3836792 | | 446.0060953 | | 442.8623508 | | 443.4702680 |
| ERROR/eV | -0.9842 | | -0.3618 | | -3.5055 | | -2.8976 |

| EXP/eV | PBE0 | scalar rel | TOTAL | | TPSS | scalar rel | TOTAL |
|---|---|---|---|---|---|---|---|
| $B^{3+}$ | -21.9943844 | -0.00642647 | -22.0008108 | | -22.0380448 | -0.0063309 | -22.0443757 |
| $B^{+}$ | -24.3019694 | -0.00706196 | -24.3090313 | | -24.3468257 | -0.00696367 | -24.3537894 |
| $B^{3+}-B^{+}$ | 62.7925865 | | 62.8098788 | | 62.8251272 | | 62.8423457 |
| ERROR/eV | -0.2928 | | -0.2755 | | -0.2603 | | -0.2431 |
| | | | | | | | |
| $C^{4+}$ | -32.3599124 | -0.01392597 | -32.3738383 | | -32.4133928 | -0.01375567 | -32.4271485 |
| $C^{2+}$ | -36.4714124 | -0.01569194 | -36.4871044 | | -36.5264308 | -0.01552389 | -36.5419547 |
| $C^{4+}-C^{2+}$ | 111.8796158 | | 111.9276705 | | 111.9214643 | | 111.9695800 |
| ERROR/eV | -0.5017 | | -0.4536 | | -0.4598 | | -0.4117 |
| | | | | | | | |
| $N^{5+}$ | -44.7252732 | -0.02655546 | -44.7518287 | | -44.7885799 | -0.02627955 | -44.8148594 |
| $N^{3+}$ | -51.142549 | -0.0305061 | -51.1730551 | | -51.2077562 | -0.0302359 | -51.2379921 |
| $N^{5+}-N^{3+}$ | 174.6229678 | | 174.7304704 | | 174.6746830 | | 174.7823410 |
| ERROR/eV | -0.7407 | | -0.6332 | | -0.6889 | | -0.5813 |
| | | | | | | | |
| $O^{6+}$ | -59.0908699 | -0.04630518 | -59.1371750 | | -59.1640138 | -0.04589044 | -59.2099043 |
| $O^{4+}$ | -68.3145916 | -0.0540794 | -68.3686710 | | -68.3899741 | -0.05367608 | -68.4436502 |
| $O^{6+}-O^{4+}$ | 250.9902524 | | 251.2017998 | | 251.0511668 | | 251.2630249 |
| ERROR/eV | -1.0275 | | -0.8160 | | -0.9666 | | -0.7548 |
| | | | | | | | |
| $F^{7+}$ | -75.4563904 | -0.07560687 | -75.5319973 | | -75.5393707 | -0.07501221 | -75.6143829 |
| $F^{5+}$ | -87.9870901 | -0.08940119 | -88.0764913 | | -88.0726945 | -0.08884967 | -88.1615442 |
| $F^{7+}-F^{5+}$ | 340.9777061 | | 341.3530689 | | 341.0491123 | | 341.4256488 |
| ERROR/eV | -1.3722 | | -0.9968 | | -1.3008 | | -0.9243 |
| | | | | | | | |
| $Ne^{8+}$ | -93.8219151 | -0.11697981 | -93.9388950 | | -93.9147348 | -0.11616031 | -94.0308951 |
| $Ne^{6+}$ | -110.15986 | -0.13964604 | -110.2995059 | | -110.255686 | -0.13893556 | -110.3946219 |
| $Ne^{8+}-Ne^{6+}$ | 444.5781202 | | 445.1948998 | | 444.6599415 | | 445.2796875 |
| ERROR/eV | -1.7898 | | -1.1730 | | -1.7080 | | -1.0882 |

| EXP/eV | TPSSh | scalar rel | TOTAL | | B97D | scalar rel | TOTAL |
|---|---|---|---|---|---|---|---|
| $B^{3+}$ | -22.0375834 | -0.00631732 | -22.0439007 | | -22.0323739 | -0.00631674 | -22.0386907 |
| $B^{+}$ | -24.3452001 | -0.00694928 | -24.3521494 | | -24.3310434 | -0.00694025 | -24.3379836 |
| $B^{3+}-B^{+}$ | 62.7934501 | | 62.8106466 | | 62.5499822 | | 62.5669486 |
| ERROR/eV | -0.2920 | | -0.2748 | | -0.5354 | | -0.5185 |
| | | | | | | | |
| $C^{4+}$ | -32.4129397 | -0.01373206 | -32.4266718 | | -32.4014329 | -0.01369666 | -32.4151296 |
| $C^{2+}$ | -36.5243465 | -0.01549803 | -36.5398445 | | -36.5002547 | -0.01543764 | -36.5156923 |
| $C^{4+}-C^{2+}$ | 111.8770771 | | 111.9251317 | | 111.5346205 | | 111.5819950 |
| ERROR/eV | -0.5042 | | -0.4562 | | -0.8467 | | -0.7993 |
| | | | | | | | |
| $N^{5+}$ | -44.7881309 | -0.02624221 | -44.8143731 | | -44.7697312 | -0.0261368 | -44.7958680 |
| $N^{3+}$ | -51.2052072 | -0.03019435 | -51.2354016 | | -51.1702439 | -0.03003646 | -51.2002804 |
| $N^{5+}-N^{3+}$ | 174.6175409 | | 174.7250841 | | 174.1668230 | | 174.2729383 |
| ERROR/eV | -0.7461 | | -0.6385 | | -1.1968 | | -1.0907 |
| | | | | | | | |
| $O^{6+}$ | -59.163572 | -0.04583389 | -59.2094059 | | -59.1378293 | -0.04562704 | -59.1834563 |
| $O^{4+}$ | -68.3869571 | -0.05361275 | -68.4405699 | | -68.3403105 | -0.05330246 | -68.3936130 |
| $O^{6+}-O^{4+}$ | 250.9810939 | | 251.1927674 | | 250.4122699 | | 250.6211285 |
| ERROR/eV | -1.0367 | | -0.8250 | | -1.6055 | | -1.3967 |
| | | | | | | | |
| $F^{7+}$ | -75.5389328 | -0.07493103 | -75.6138639 | | -75.5054765 | -0.07457674 | -75.5800532 |
| $F^{5+}$ | -88.0692026 | -0.08875567 | -88.1579582 | | -88.0101955 | -0.08821009 | -88.0984056 |
| $F^{7+}-F^{5+}$ | 340.9660066 | | 341.3421943 | | 340.2707353 | | 340.6417176 |
| ERROR/eV | -1.3839 | | -1.0077 | | -2.0792 | | -1.7082 |
| | | | | | | | |
| $Ne^{8+}$ | -93.9143 | -0.11604836 | -94.0303483 | | -93.8728341 | -0.11549736 | -93.9883315 |
| $Ne^{6+}$ | -110.251718 | -0.13880171 | -110.3905201 | | -110.1798 | -0.13793635 | -110.3177366 |
| $Ne^{8+}-Ne^{6+}$ | 444.5637989 | | 445.1829489 | | 443.7351508 | | 444.3457469 |
| ERROR/eV | -1.8041 | | -1.1850 | | -2.6327 | | -2.0222 |

| EXP/eV | BHLYP | scalar rel | TOTAL | | BP86 | scalar rel | TOTAL |
|---|---|---|---|---|---|---|---|
| $B^{3+}$ | -22.0276091 | -0.00641563 | -22.0340247 | | -22.0243344 | -0.00655152 | -22.0308859 |
| $B^{+}$ | -24.3397998 | -0.00705104 | -24.3468508 | | -24.3317254 | -0.00718613 | -24.3389116 |
| $B^{3+}-B^{+}$ | 62.9179127 | | 62.9352031 | | 62.7873077 | | 62.8045765 |
| ERROR/eV | -0.1675 | | -0.1502 | | -0.2981 | | -0.2808 |
| | | | | | | | |
| $C^{4+}$ | -32.4001037 | -0.01389677 | -32.4140005 | | -32.3991183 | -0.01415056 | -32.4132688 |
| $C^{2+}$ | -36.5179398 | -0.01566036 | -36.5336002 | | -36.508742 | -0.01591624 | -36.5246582 |
| $C^{4+}-C^{2+}$ | 112.0520269 | | 112.1000164 | | 111.8285577 | | 111.8766043 |
| ERROR/eV | -0.3293 | | -0.2813 | | -0.5527 | | -0.5047 |
| | | | | | | | |
| $N^{5+}$ | -44.7724005 | -0.02648899 | -44.7988895 | | -44.7743781 | -0.02691719 | -44.8012953 |
| $N^{3+}$ | -51.1966885 | -0.03044316 | -51.2271316 | | -51.1887105 | -0.03087021 | -51.2195807 |
| $N^{5+}-N^{3+}$ | 174.8137800 | | 174.9213785 | | 174.5428755 | | 174.6504427 |
| ERROR/eV | -0.5499 | | -0.4423 | | -0.8208 | | -0.7132 |
| | | | | | | | |
| $O^{6+}$ | -59.1449396 | -0.04619175 | -59.1911313 | | -59.1502303 | -0.04686016 | -59.1970904 |
| $O^{4+}$ | -68.3756488 | -0.05396806 | -68.4296168 | | -68.370516 | -0.05464451 | -68.4251606 |
| $O^{6+}-O^{4+}$ | 251.1803913 | | 251.3919954 | | 250.8967556 | | 251.1085784 |
| ERROR/eV | -0.8374 | | -0.6258 | | -1.1210 | | -0.9092 |
| | | | | | | | |
| $F^{7+}$ | -75.5174092 | -0.07542349 | -75.5928327 | | -75.5262242 | -0.07641098 | -75.6026352 |
| $F^{5+}$ | -88.0545934 | -0.08921624 | -88.1438096 | | -88.0534236 | -0.09022573 | -88.1436494 |
| $F^{7+}-F^{5+}$ | 341.1541569 | | 341.5294766 | | 340.8824590 | | 341.2583774 |
| ERROR/eV | -1.1957 | | -0.8204 | | -1.4674 | | -1.0915 |
| | | | | | | | |
| $Ne^{8+}$ | -93.889897 | -0.11670896 | -94.0066060 | | -93.9023175 | -0.11810071 | -94.0204183 |
| $Ne^{6+}$ | -110.233482 | -0.13937373 | -110.3728553 | | -110.237073 | -0.14080769 | -110.3778806 |
| $Ne^{8+}-Ne^{6+}$ | 444.7315878 | | 445.3483277 | | 444.4913342 | | 445.1092225 |
| ERROR/eV | -1.6363 | | -1.0196 | | -1.8766 | | -1.2587 |